# Improvement of the $Tm^{3+}$:$^3H_4$ level lifetime in silica optical fibres by lowering the local phonon energy

B. Faure [a], W. Blanc [a], B. Dussardier [a,*], G. Monnom [a]

[a] Laboratoire de Physique de la Matière Condensée

Université de Nice Sophia-Antipolis - CNRS, UMR 6622

Parc Valrose - F-06108 NICE CEDEX 2 – France

* Corresponding author, Tél: +33 492 076 748, Fax: +33 492 076 754,

email address: bernard.dussardier@unice.fr

## Abstract

The role of some glass network modifiers on the quantum efficiency of the near-infrared fluorescence from the $^3H_4$ level of $Tm^{3+}$ ion in silica-based doped fibres is studied. Modifications of the core composition affect the spectroscopic properties of $Tm^{3+}$ ion. Adding 17.4 mol% of $AlO_{3/2}$ to the core glass caused an increase of the $^3H_4$ level lifetime up to 50 µs, 3.6 times higher than in pure silica glass. The quantum efficiency was increased from 2% to approximately 8%. On the opposite, 8 mol% of $PO_{5/2}$ in the core glass made the lifetime decrease downto 9 µs. These changes of $Tm^{3+}$ optical properties are assigned to the change of the local phonon energy to which they are submitted by modifiers located in the vicinity of the doping sites. Some qualitative predictions of the maximum achievable quantum efficiency are possible using a simple microscopic model to calculate the non-radiative de-excitation rates.

**Introduction**

Thulium doped fibres have been widely studied in the past few years. Because of $Tm^{3+}$ ion rich energy diagram, lasing action and amplification at multiple infrared and visible wavelengths are allowed. Thanks to the possible stimulated emission peaking at 1.47 µm ($^3H_4$ ⇔ $^3F_4$, see Fig. 1), discovered by Antipenko *et al.* [1], one of the most exciting possibilities of $Tm^{3+}$ ion is amplifying optical signal in the S-band (1.47-1.52 µm), in order to increase the available bandwidth for future optical communications. Unfortunately, the upper $^3H_4$ level of this transition is very close to the next lower $^3H_5$ level so non-radiative de-excitations (NRD) are likely to happen in high phonon energy glass host, causing detrimental gain quenching.

In standard silica based optical fibres the maximum phonon energy is high: $E_p$ ~ 1100 cm$^{-1}$. Then only three vibrational quanta (phonons) are needed to bridge the energy gap between the $^3H_4$ and $^3H_5$ multiplets ($\Delta E$ ~3700 cm$^{-1}$). Consequently the measured $^3H_4$ level lifetime $\tau$ is 14 µs only [2], much shorter than the purely radiative lifetime $\tau_{rad}$ = 650 µs [3], and the quantum efficiency is only ~2%. Therefore efficient host materials for implementing $Tm^{3+}$ ion in the S-band have been restricted to non-silicate glasses with low phonon energy. In fluoride glass host ($E_p$ ~ 580 cm$^{-1}$) five phonons are needed to bridge $\Delta E$, the probability of NRD is very low and the de-excitation is almost purely radiative. As a consequence, the $^3H_4$ level lifetime almost equals the radiative lifetime, $\tau$ ~ $\tau_{rad}$ = 1.3 ms [3] and the quantum efficiency approaches 100%. This is why the first Thulium Doped Fibre Amplifier (TDFA) was realized in 1995 by Komukai *et al.* in fluoride glass [4]. There has been many investigations for optimizing pumping schemes with newly available pump sources [5,6]. Recently, TDFAs made of other low phonon glasses have been investigated. In tellurite glass ($E_p \sim 700$ cm$^{-1}$) amplification in the S-band is possible thanks to the 90% quantum efficiency and the 310-µs lifetime [7,8]. TDFAs using multicomponent silicate glasses were also realized [9,10] but small information about the glass composition and preparation of the samples was given in the cited references. The potential of alkaline-earth aluminate glass ($E_p \sim 780$ cm$^{-1}$) for TDFA was studied [11]: S-band quantum efficiency of 35% and $^3H_4$ level lifetime of around 230 µs were reported. Note that using low phonon energy matrix like oxyfluoride glass ceramic ($E_p \sim 300$ cm$^{-1}$) is possible [12,13]. Unless they show good amplification results, all those fibre glasses or ceramics can not be easily used in telecommunication networks based on silica fibres, because of their poor chemical durability, high production price, low reliability. As they are not fusion-spliceable to silica fibres one must use other techniques such as connectors or glued butt-coupling that are more lossy and easily damaged under high pumping level. For those reasons, it would still be very interesting to demonstrate a TDFA based on silica glass.

To address this problem, we propose to study the effect of some modifications of Tm$^{3+}$ ion local environment. Keeping the overall fibre composition as close as possible to that of a standard silica fibre, we expect to control the rare earth spectroscopic properties by codoping with selected modifying oxides. We have studied the incorporation of modifying elements compatible with a standard fabrication technique, such as modified chemical vapour deposition (MCVD). GeO$_2$ is widely used in silica fibre fabrication to increase the core refractive index. AlO$_{3/2}$ is a promising candidate because it is known to improve the spectroscopic properties of Er$^{3+}$ ion for C-band amplification [14]. It is also used to reduce quenching effect through clustering in highly rare earth-doped silica [15]. Both oxides have a lower maximum phonon energy than silica. We use high phonon energy $PO_{5/2}$ as opposite demonstration.

In this article, we describe the samples preparation and compositional characterization, then the absorption, fluorescence and fluorescence decay measurement techniques. We use measured lifetimes to derive the quantum efficiency of $Tm^{3+}$ ion. Then we present the results and discuss the effects of the core composition on the spectroscopic properties of thulium in the fibres. In particular, we show that co-doping with phosphorus or aluminum results in the strong modification of the local environment of thulium, whereas germanium codoping has a lesser impact. Aluminum is the most interesting modifier in the series studied here, as a four-times increase in the $Tm^{3+}$ ion quantum efficiency is observed when codoping with less than 10 mol% of $Al_2O_3$. These results are in qualitative agreement with a simple model implementing multiphonon NRD empirical theory, which shows that the striking differences between codopants lie in the local phonon energy alteration they induce to $Tm^{3+}$ ion.

## 1. Experimental

### 1.1 *Sample preparation*

The $Tm^{3+}$-doped optical fibres were drawn from silica-based preforms using a drawing tower. The preform samples were prepared from pure silica tube by classical MCVD method allowing phosphorus and/or germanium to be incorporated in the glass [16]. Doping with thulium and/or aluminium was performed using the so called "solution doping technique" [17]. The core silica layer was deposited at lower temperature than the preceding cladding layers, so that it left porous. Then the substrate tube was filled with an alcoholic solution of $TmCl_3·6H_2O$ and/or $AlCl_3·6H_2O$ salts to impregnate the porous layer. After few hours the solution was removed, the porous layer was dried and sintered, and the tube was collapsed into a 10 mm-diameter solid preform. We have made three types of $Tm^{3+}$-doped preforms and fibres: *Tm(Al)* codoped with various contents of aluminium, *Tm(Ge)* with germanium and *Tm(P)* with phosphorus. Germanium, aluminium and phosphorus were also used to raise the refractive index of the core. Note that for manufacturing purpose, all samples contained a small quantity of phosphorus in the core (< 1 mol% of $PO_{5/2}$), except for the *Tm(P)* sample which contained 8 mol% (see Table 1). We will see that the effect of this low P codoping is negligible. In order to avoid clustering effects, such as energy transfer between rare earth ions, $Tm^{3+}$ ion concentration was kept below 200 mol ppm.

### 1.2  *Core composition*

The oxide core composition of each sample was deduced from the refractive index difference (*Δn*) between core and cladding in the preform, knowing the correspondence between index rising and $AlO_{3/2}$, $GeO_2$, $PO_{5/2}$ concentration in silica glass [18,19]. *Δn* was measured with a commercial refractive index profile analyser. Note that the index rising due to $Tm^{3+}$ ion was neglected because of the very small concentration. The composition was directly measured on some preforms using electron probe microanalysis technique in order to compare the results: a good agreement was found. All results are summarized in Table 1. The $Tm^{3+}$ ion concentration was deduced from the 785 nm ($^3H_6 => ^3H_4$) absorption peak measured in fibres and using the absorption cross section reported in [2]: $\sigma_{abs}(785\ nm) = 8.7 \times 10^{-25}\ m^2$.

### 1.3  *Absorption measurements*

Ground state absorption (GSA) spectra were also measured from fibres to study the spectroscopic variations versus codopant type and concentration, and evaluate *ΔE* the energy gap between the $^3H_5$ and $^3H_4$ levels of the rare earth in modified silica environment (Fig. 1). We used a classical cut-back method which consists of measuring two different transmission spectra for two lengths of the tested fibre. We used a halogen lamp coupled with a 2 nm resolution monochromator. The output infrared light was registered with either a Ge or Si detector. The white light beam was modulated and the detector signal analyzed by a SR-830 DSP lock-in amplifier. Low temperature (77K) measurements were performed by dipping the tested fibre connected to transparents fibre patch-cords into liquid nitrogen. A typical GSA spectrum is shown in Fig. 2.

### 1.4  *Fluorescence decay measurements*

The role of the core composition on the NRD of $Tm^{3+}$ ions from the $^3H_4$ level was investigated by measuring the lifetime of the 810-nm fluorescence from the $^3H_4$ level downto the $^3H_6$ ground-level. The beam from a modulated 785-nm, 20-mW fibre-coupled laser diode was coupled into the tested fibre through a single-mode fibre-coupler. The pump wavelength was tuned to that of the $^3H_6 => ^3H_4$ absorption peak. The coupler transmitted 60% of the pump to the Tm-doped fibre. To avoid spectral distorsion caused by re-absorption of the signal, the 810-nm fluorescence was measured counterpropagatively. Although a weak residual absorption exists, the setup and procedure were chosen to minimize its impact. The fluorescence was collected from the second coupler arm and 48% of it was directed to an optical spectrum analyser (Anritsu MS9030A-MS9701B) equiped with a –20 dB optical through-put port. This port was tuned to 810 nm (1 nm spectral width). The output light was detected by an amplified avalanche silicon photodiode (APD) (EG&G SPCM AQR-14-FC) operated in the photon-counting mode. The TTL electrical pulses from the APD were counted by a Stanford SR400 photon counter synchronized by the laser diode modulation signal. Decay curves were registered using a time-gate scanning across one pump modulation period. In order to minimize errors (laser fluctuations, …) and increase the S/N ratio, the signal was normalized in real time by the signal from a fixed time-gate integrating the signal over a full

500 μs-modulation period, and each data point was averaged 1000 times. A decay curve contained typically 800 data points.

## 2. Results

### 2.1 *Effect of core composition on absorption*

The room temperature GSA spectra from the fondamental $Tm^{3+}:^3H_6$ multiplet of *Tm(Al)*, *Tm(Ge)* and *Tm(P)* fibre samples are shown in Fig. 3. The GSA to the $^3H_4$ level is shown in Fig. 3(a). The *Tm(Al)* samples $^3H_4$ spectra are very similar, except for small changes on the wings of the band. For the seek of clarity we show only one GSA spectrum from the *Tm(Al)* sample series. More importantly the $^3H_4$ peak wavelength is the same for all *Tm(Al)* samples, well within the spectral resolution of the measurement (2 nm), and even when measured at 77 K (not shown here). As no energy level shift nor broadening was observed, we conclude that varying the aluminum content in the core has almost no effect on the $Tm^{3+}$ ion local crystal field intensity. Fig. 3(b) shows the GSA to the $^3H_5$ multiplet around 1200 nm from *Tm(Al)*, *Tm(P)*, and *Tm(Ge)* samples. In this figure, the curves are normalized to the height of the characteristic shoulder around 1175 nm. This is done so because the highest peak at ~1210 nm has an hypersensitive character depending of the material composition, whereas the other $^3H_5$ band features are much less sensitive.

Both $^3H_4$ and $^3H_5$ *Tm(Al)* GSA spectra have broader and smoother features than the spectra from the *Tm(Ge)* and *Tm(P)* samples. This is characteristic of aluminum codoping in rare-earth-doped silica fibres and is mostly attributed to inhomogeneous broadening, whereas phosphorus reduces inhomogeneous broadening compared to silica [14]. The *Tm(P)* sample exhibits a red shift of the mean wavelength on both bands, whereas the *Tm(Ge)* sample exhibits a blue shift only visible on the $^3H_4$ band. These spectra will be used below to determine *ΔE* for each sample type.

## 2.2 *Effect of core composition on the fluorescence lifetime*

Fig. 4 shows the normalized decay curves of the 810-nm fluorescence from the $^3H_4$ level, after the pump was turned off, from *Tm(Al)* fibre samples with increasing aluminum content in the core. It is worth noting that all decay curves are non-exponential, as reported before [20]. This can be attributed to several phenomena. A first cause could be inter-ionic energy transfer during relaxation. Recently D. Simpson showed that the ($^3F_4 => ^3H_4$ : $^3F_4 => ^3H_6$) energy transfer happens in silica fibres even at low $Tm^{3+}$ concentrations (~30 ppm) when the $^3F_4$ level is resonantly excited [21]. However he showed that there was no energy transfer when the $^3H_4$ level is resonantly excited, as we did for the present study. Therefore, any effect from $Tm^{3+}$-$Tm^{3+}$ energy transfer can be ruled out.

A prefered interpretation follows Lincoln's tentative attribution to the inhomogeneous broadening and perhaps low homogeneous broadening of the $Tm^{3+}$ ion atomic levels in silica [20]. The $Tm^{3+}$ ions are dispersed as 'ions packets' in a quasi-continuous distribution of sites in the matrix [22]. It was proposed that the $^3H_4$ level was very sensitive to even slight local variations in composition or coordination. Then spectroscopic characteristics including central emission wavelengths and lifetimes are described by continuous distribution functions. Hence the resulting decay curves from all ion-packets would be a superposition of simple exponential decays. This was shown for other dopants in silicate glasses, such as $Cr^{3+}$ ion [23]. Although this effect is more dramatic in transition metal doped materials, because their 3d orbitals are not screened like the rare-earth ions 4f orbitals, the interpretation is still valid here. We have reported elsewhere on a continuous decay analysis applied to $Tm^{3+}$ ions in silica fibres [24], derived from that developed for $Cr^{3+}$ in glasses and which conforts this interpretation. In the scope of this article, we study the variations of 1/e-lifetimes ($\tau$) versus the concentration in network modifier oxides (Fig. 5). The lifetime value for $Tm^{3+}$-doped pure silica is taken from reference [2]. The lifetime strongly changes with the composition of the glass host. The most striking results are observed within the Tm(Al) sample series: $\tau$ linearly increases with increasing $AlO_{3/2}$ content, from 14 μs in pure silica to 50 μs in sample *Tm(Al)-6* containing 17.4 mol% of $AlO_{3/2}$. The lifetime was increased about 3.6 times. Fibres with higher contents in aluminum were prepared, however scattering losses appeared in the fibres. The lifetime of the 20 mol% $GeO_2$ doped fibre *Tm(Ge)* was increased upto 28 μs whereas that of the 8 mol% $PO_{5/2}$ doped fibre *Tm(P)* was reduced downto 9 μs. We see that aluminum codoping seems the most interesting route among the three tested codopants. The potential improvement of the amplification in the S-band of aluminium-doped silica-based TDFA has been studied elsewhere [25,26], and is out of the scope of this paper.

### 2.3  *Effect of core composition on non-radiative de-excitation (NRD)*

The measured lifetime $\tau$ depends on the radiative lifetime $\tau_{rad}$, and the NRD rate $W_{nrd}$:

$$1/\tau = 1/\tau_{rad} + W_{nrd} \qquad (1)$$

The value of $\tau_{rad}$ is 650 μs in silica glass as calculated from the phenomenological Judd-Ofelt model [3]. The NRD rate is given by the empirical formula [27]:

$$W_{nrd} = W_0 \exp\{-\alpha(\Delta E - 2 E_p)\} \qquad (2)$$

where $W_0$ and $\alpha$ are constants weakly depending on the glass host. Within various oxide compounds, these constants have similar values (see Table 3) [28]. $\Delta E$ is the energy gap between the $^3H_4$ and $^3H_5$ multiplets. $E_p$ is the maximum phonon energy of the host. Two phenoma affect the NRD rate: (i) modification of the crystalline field applied on $Tm^{3+}$ ion, hence altering the energy levels and thus $\Delta E$, and/or (ii) modification of the maximum effective phonon energy to which the rare earth is submitted when one or several network modifiers are in the neighbourhood. In the following we will study and discuss the effect of crystal field changes within samples and show that this effect is negligible compared to phonon local energy variations.

First we establish a good estimation of the energy gap. $\Delta E$ is measured between the lowest Stark level from the $^3H_4$ multiplet and the highest Stark level from the $^3H_5$ multiplet (Fig. 6(a)). This gap is bridged via multi-phonons de-excitations in order to cause NRD. However because of the inhomogeneous and homogeneous spectral broadening, the absorption bands show finite width and smooth features even at low temperature, rendering difficult the determination of the energy of the relevant Stark levels, and therefore of $\Delta E$. Its value is empirically estimated by considering the lifetime measured in pure silica (14 µs). This lifetime can be obtained by adjusting $\Delta E$ to 3693 cm$^{-1}$ (rounded to 3700 cm$^{-1}$ in the following) in equation (2) and using parameters for silica from Table 3. This value in pure silica is compared with estimations based on experimental GSA measurements. Fig. 6(b) shows the case of one particular *Tm(Al)* sample. We estimated $\Delta E$ as follows: the most probable (and intense) Stark-Stark transition on the Tm$^{3+}$:$^3H_6$-$^3H_4$ GSA band is between the lowest Stark levels in each multiplet, corresponding to the peak of the GSA spectrum [20]. The energy (resp. wavelength) of the lowest $^3H_4$ Stark level is estimated at $E(^3H_{4\text{-lowestStark}})$ = 12720 cm$^{-1}$ (786 nm) in *Tm(Al)* samples. On the other hand, the highest $^3H_5$-Stark level causes the highest energy absorption contribution to the $^3H_6$-$^3H_5$ band on the spectrum. As this contribution is weak and smooth, even in 77K absorption curves, we arbitrarily take the energy at which the $^3H_5$ absorption band is 10% relative to the height of the shoulder peaking around 8500 cm$^{-1}$ (1175 nm), on the low energy wing (see Fig. 6(b)). This energy is $E(^3H_{5\text{-highestStark}})$ ~ 9090 cm$^{-1}$. From this measurement $\Delta E$ = 3630 cm$^{-1}$. The same procedure is applied to all samples (Table 2). Note that all $\Delta E$ values lie within ±2% compared to that estimated from pure silica, whereas accuracy is ~ ±1%.

## 3. Discussion

From this point, we calculate $W_{NRD}$ along two possible microscopic models of which we will discuss the validity. In the first model it is assumed that the glass modifiers are homogeneously dispersed in the glass. As all $Tm^{3+}$ ions sustain the same local vibrational environment, the local environment is equivalent to pure silica in terms of phonon transfer and propagation properties. Variations within sample types are only characterized by $\Delta E$ changes. In this case, $W_{NRD}$ and $\tau_{calc}$ are calculated using equations (2) and (1), respectively, using $E_p$, $W_0$ and $\alpha$ from pure silica [28,29], $\Delta E$ found for the three types of samples (Table 3) and $\tau_{rad}$ = 650 µs. The calculated lifetimes along the homogeneous model for $AlO_{3/2}$, $GeO_2$ and $PO_{5/2}$ are 10, 20 and 17 µs, respectively. These values do not fit with our observations on *Tm(Al)* and *Tm(P)* samples and give opposite tendencies, whereas *Tm(Ge)* calculated value is lower than the measured 28 µs value (Fig. 5). We conclude that the lifetime variations within the studied samples cannot be explained only by small variations of $\Delta E$ and that the homogeneous model is not applicable here.

We consider here a second model (called the *inhomogeneous model*) already proposed for rare-earth-doped alumino- and phospho-silicate glasses either prepared by standard techniques [15,30], by MCVD and solution-doped [31] and also by sol-gel [32]. Aluminum and phosphorus have medium and strong affinities with the rare-earth ion, respectively, and are precisely located within its second coordination sphere. Saitoh *et al.* interpret some optical spectroscopic properties using this model. Here we propose that phonon coupling, and therefore *NRD*'s probabilities are mostly influenced by phonon properties from the modifier oxide. Therefore in equation (2) we use the effective phonon energy $E_p$ of an hypothetic pure modifier oxide. $W_{NRD}$ and $\tau_{calc}$ are calculated using equations (2) and (1), using $E_p$, $W_0$ and $\alpha$ from the three respective modifier oxides and $\Delta E$ from the three types of samples. Note that because the $W_0$ and $\alpha$ values for $AlO_{3/2}$ were not known, we arbitrarily used average values calculated from a series of oxides reported in [28,29](data in italics in Table 3). As the radiative lifetime $\tau_{rad}$ was not available for all types of glasses we assumed $\tau_{rad}$ = 650 μs as for silica. This assumption is justified because (i) the $Er^{3+}$ overall transition strength shows small variations across a wide range of oxide glasses and (ii) even a substantial error in $\tau_{rad}$ estimation has a small impact on calculating the lifetime $\tau$ from equation (1).

It is evident that $\tau_{calc}$ reported in Table 3 are in better agreement with our observations, compared to those from the *homogeneous* model. Phosphorus co-doping greatly lowers the lifetime, whereas aluminum codoping increases it by almost an order of magnitude. Indeed phosphorus in silica brings energetic P=O double bonds and has a strong affinity to the rare-earth ion, causing a high $W_{NRD}$ probability. Germanium is a network former substituted to silicon. Therefore it is more likely homogeneously distributed within the silica network, has a weak interaction with rare-earth ions and a weak influence on *NRD*s. Although $GeO_2$ phonon energy is lower than that of silica, increasing Ge content lowers $W_{NRD}$ to a lesser extent than for aluminum codoping. This explains the large discrepancy between measured and calculted lifetimes for the *Tm(Ge)* sample.

In the *Tm(Al)* series our previous empirical investigations on the non-exponential nature of the decay curves from these samples [24] have shown the presence of two distinct families of $Tm^{3+}$ sites: the first was characterized by a short lifetime (~10 μs) and was attributed to Al-poor sites, whereas the second had longer lifetimes (~100 μs) and was attributed to Al-rich sites. The calculated lifetimes for the inhomogeneous model in *Tm(Al)* samples confirm our findings. It is proposed that increasing the average aluminum content increases the probability that a $Tm^{3+}$ ions be located in an Al-rich site, therefore contributing to increasing the lifetime and lowering the non-exponentiality of the decay curves. Although all samples contained a small amount of phosphorus, only the characteristics brought by aluminum-codoping had a noticeable effect on the lifetime. The model indicates that a maximum ~100 μs lifetime is expected from aluminum codoping in silica fibres.

## 4. Conclusion

We have studied the effect of modifying the composition of $Tm^{3+}$-doped silica-based optical fibres on the $^3H_4$ level lifetime. Al and P doping are efficient to increase and decrease the lifetime and quantum efficiency, respectively. An addition of 17.4 mol% of $AlO_{3/2}$ to silica glass increases the thulium $^3H_4$ level lifetime by 3.6 to above 50 μs, whereas 8 mol% of $PO_{5/2}$ reduces it to 9 μs only. Ge doping has a benefic effect, but is less efficient than Al. This behaviour versus core composition is attributed to the strong alteration of the multiphonon NRD rate. We have shown that the NRD rate is very sensitive to the local phonon energy, dictated by the presence of modifiers oxides in the direct vicinity of the $Tm^{3+}$ ions. A simple model of multiphonon NRD using the maximum phonon energy $E_p$ of hypothetic pure modifying oxides, such as $AlO_{3/2}$ or $PO_{5/2}$, give good qualitative agreement with experimental results. This allows to predict trends toward increasing the quantum efficiency of emission bands from the $^3H_4$ level of the $Tm^{3+}$ ion, either by increasing the concentration in aluminum or selecting more appropriate modifiers among heavier elements that can be incorporated using MCVD or other commercial technique.

**Acknowledgements**

BF acknowledges support from the French Ministry of Research and Technology for his PhD scholarship. Laboratoire de Physique de la Matière Condensée is part of the CNRS-GIS *GRIFON* multisite optical fibre technological consortium. Electron probe microanalysis measurements were performed at Institute of Photonics and Electronics, Academy of Sciences of the Czech Republic, Prague.

**Figure Captions**

Fig. 1 : Schematic energy diagram of $Tm^{3+}$ ion, showing the relevant multiplets. Solid arrows: absorption and emission optical transitions; thick arrow: NRD (non-radiative de-excitation) across the energy gap between the $^3H_4$ and $^3H_5$ multiplets, $\Delta E$~ 3700 cm$^{-1}$.

Fig. 2: Normalized absorption (dash) and emission (solid) spectra of the $Tm^{3+}$: $^3H_6 \Leftrightarrow {}^3H_4$ transition in a silica fibre.

Fig. 3: Normalized GSA spectra of the (a) $^3H_6 => {}^3H_4$ and (b) $^3H_6 => {}^3H_5$ transitions in the *Tm(Al)*, *Tm(Ge)* and *Tm(P)* samples, respectively. Curves in (b) are normalized to the height of the shoulder around 1175 nm (see text).

Fig. 4: Aluminium doped sample decay curves of the 810 nm fluorescence in semi-logarithmic scale.

Fig. 5: $^3H_4$-level lifetime (e$^{-1}$) from *Tm(Al)* (filled squares), *Tm(Ge)* (circle) and *Tm(P)* (triangle) samples.

Fig. 6 : (a) Schematic manifold level of $Tm^{3+}$ ions. (b) GSA spectrum from $^3H_6 => {}^3H_5$ and $^3H_4$ in a *Tm(Al)* fibre sample.

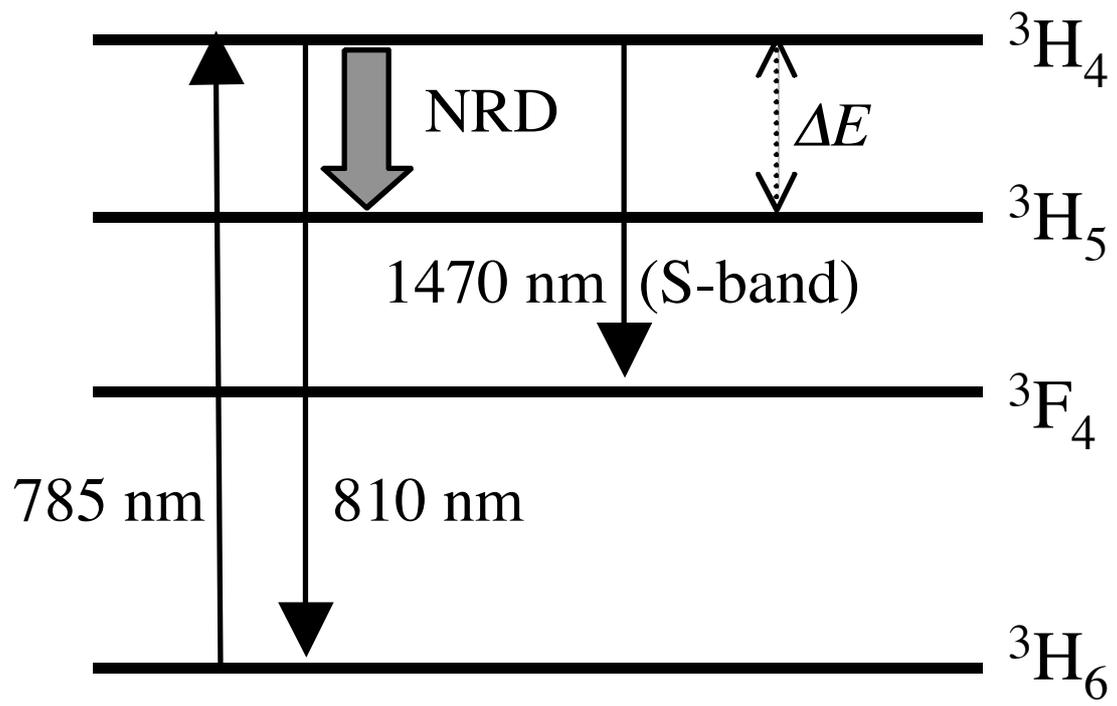

Fig. 1

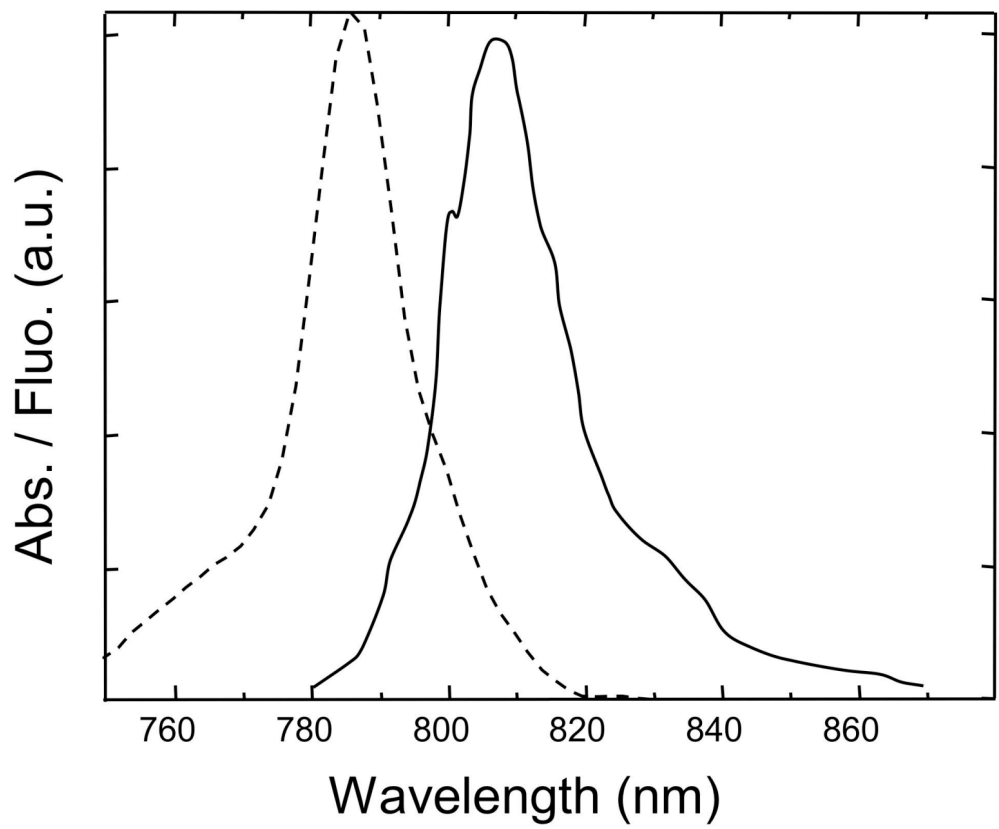

Fig. 2

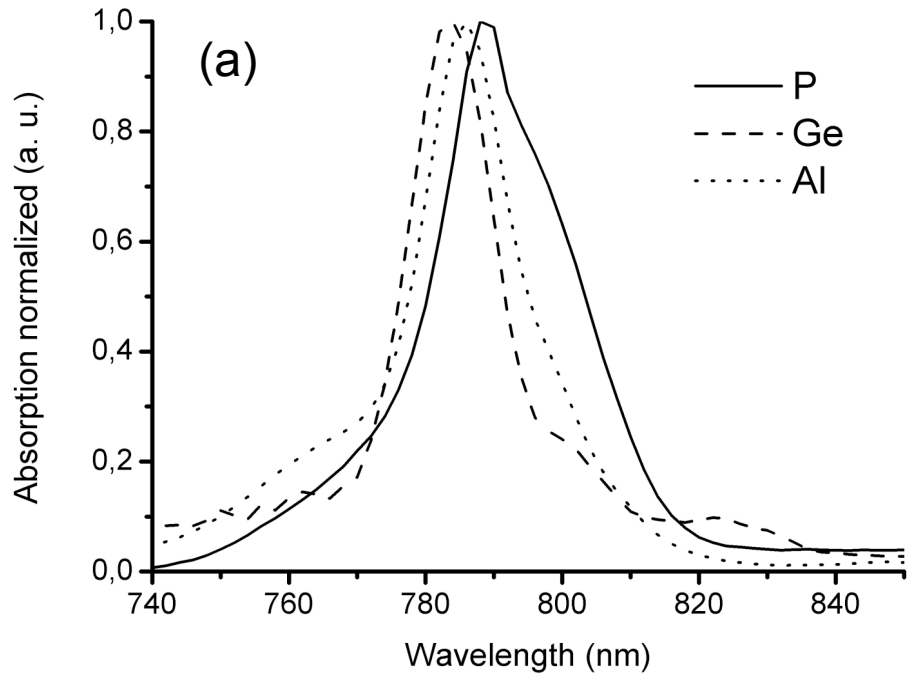

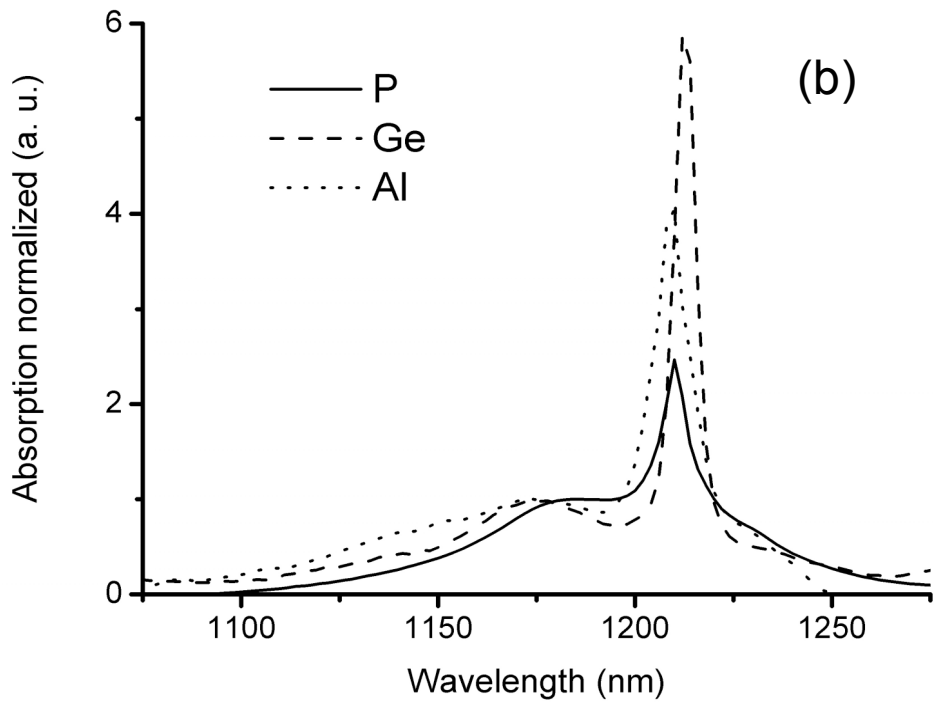

Fig. 3

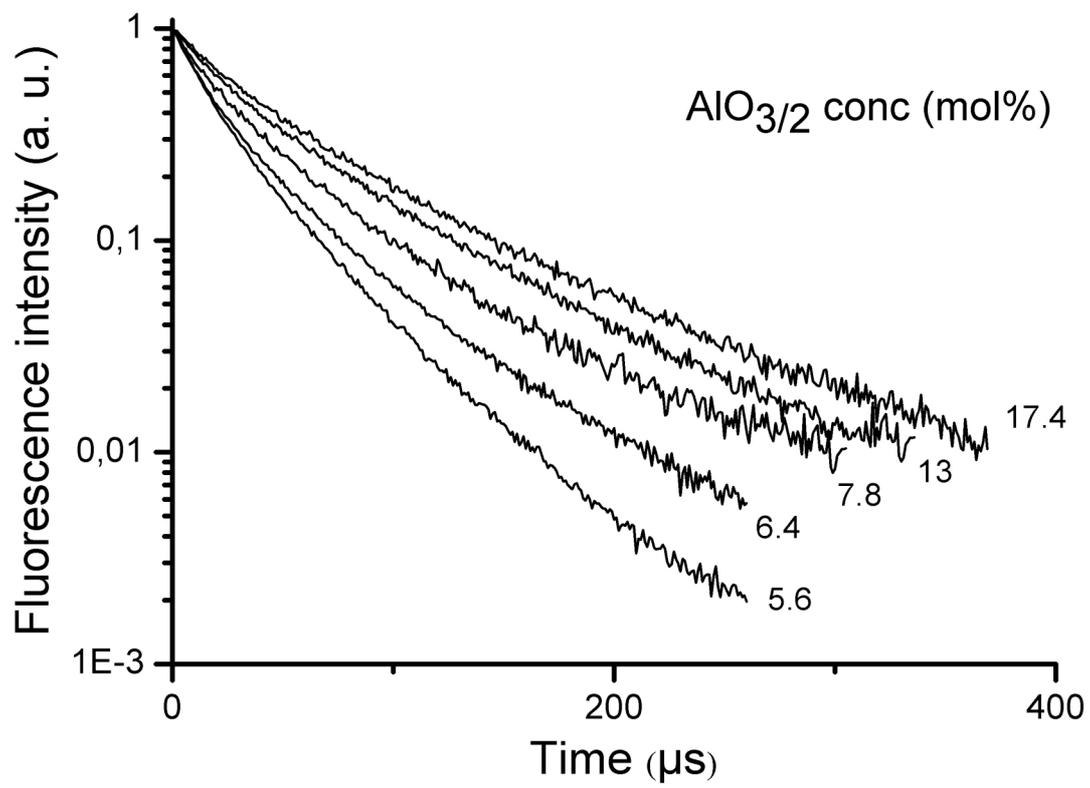

Fig. 4

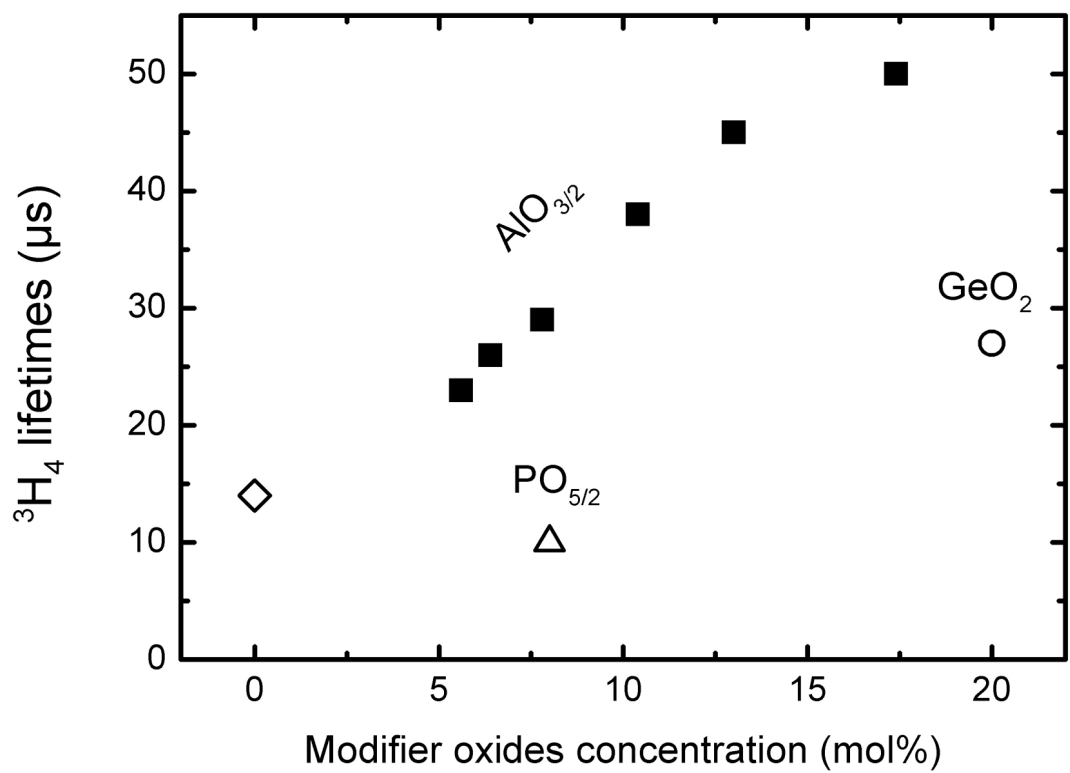

Fig. 5

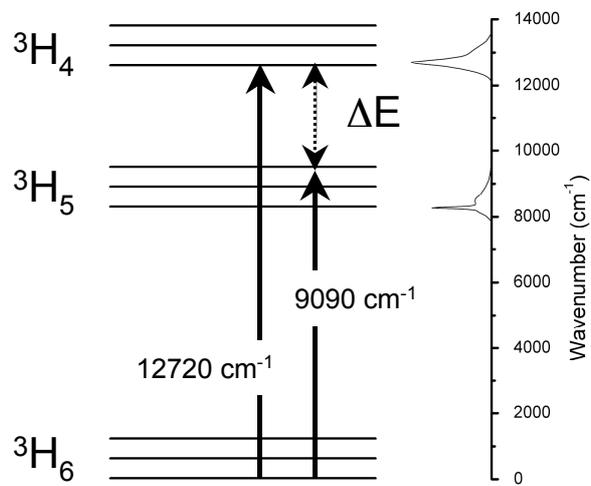

(a)

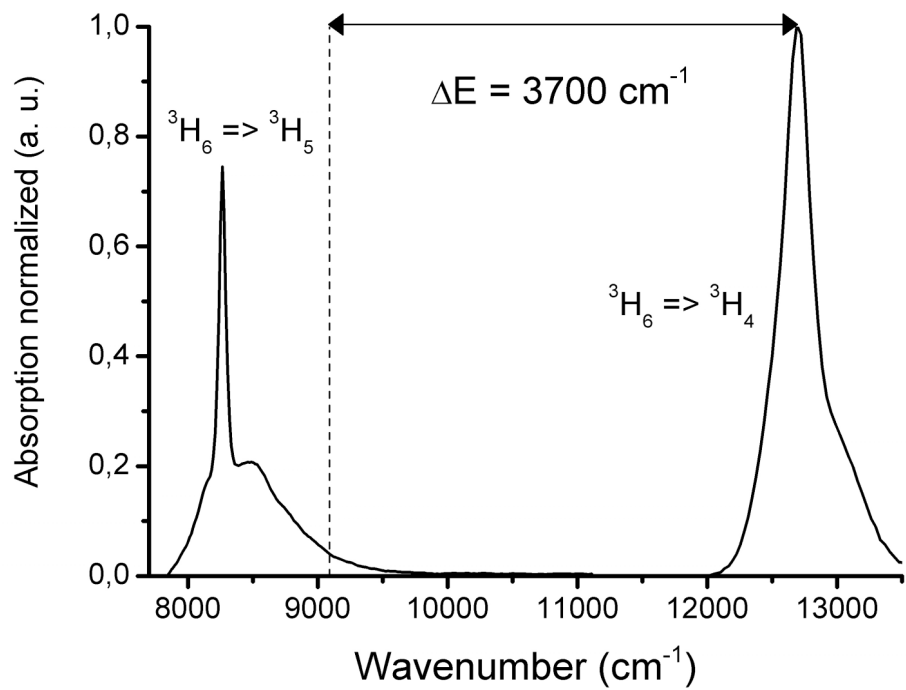

(b)

Fig. 6

# Tables

Table 1

Core composition of the $Tm^{3+}$-doped samples. Uncertainty ± 0.5 mol%

|  | modifier | $Tm^{3+}$ |
|---|---|---|
|  | (mol %) | (mol ppm) |
| *Tm(Ge)* | $GeO_2$ |  |
|  | 20 | 70 |
| *Tm(P)* | $PO_{5/2}$ |  |
|  | 8 | 200 |
| *Tm(Al)-#* | $AlO_{3/2}$ |  |
| -1 | 5.6 | 30 |
| -2 | 6.4 | 40 |
| -3 | 7.8 | 60 |
| -4 | 10.4 | 80 |
| -5 | 13.0 | 30 |
| -6 | 17.4 | 40 |

Table 2

Energy of the $^3H_4$ lowest Stark level and the $^3H_5$ highest Stark level in *Tm(X)* samples, *X=Al, Ge* or *P*, and *ΔE* estimated values.

|  | $^3H_6 =>^3H_4$ peak (cm$^{-1}$) | $^3H_6 =>^3H_5$ edge 10% (cm$^{-1}$) | *ΔE* (cm$^{-1}$) |
|---|---|---|---|
| *Tm(Al)* | 12710 | 9090 | 3620 |
| *Tm(P)* | 12670 | 8940 | 3730 |
| *Tm(Ge)* | 12760 | 8990 | 3770 |
| max.error | ± 20 | ±20 | ±40 |

Table 3

NRD parameters [28,29] and calculated NRD rates and lifetimes for hypothetic pure Si, Al, Ge and P oxides, respectively, with $\tau_{rad}$ = 650 µs. Errors on *ΔE*(~1 %, see Table 2) induce ~ ±20 % errors on $W_{nrd}$ and $\tau_{calc}$.

| Oxide | $E_p$ (cm$^{-1}$) | $W_0$ (10$^7$ s$^{-1}$) | $\alpha$ (10$^{-3}$ cm) | $W_{NRD}$ (s$^{-1}$) | $\tau_{calc}$ (µs) |
|---|---|---|---|---|---|
| SiO$_2$ | 1100 | 7.8 | 4.7 | 67658 | 14 |
| AlO$_{3/2}$ | 870 | *6.2* | *4.7* | 8206 | 102 |
| GeO$_2$ | 900 | 6.1 | 4.6 | 7075 | 116 |
| PO$_{5/2}$ | 1320 | 7.6 | 4.7 | 452817 | 2 |